\documentclass[aps,pre,twocolumn,groupedaddress,showpacs]{revtex4}
\usepackage{graphicx}

\newcommand{\comment}[1]{}

\begin{document}

\title{Transient and stationary behavior of the
Olami-Feder-Christensen earthquake model}


\author{Felix Wissel and Barbara Drossel}
\affiliation{Institut f\"ur Festk\"orperphysik,  TU Darmstadt,
Hochschulstra\ss e 6, 64289 Darmstadt, Germany }

\date{\today}

\begin{abstract}
Using long-term computer simulations and mean-field like arguments, we
investigate the transient time and the properties of the stationary
state of the Olami-Feder-Christensen earthquake model as function of
the coupling parameter $\alpha$ and the system size $N$. The most
important findings are that the transient time diverges
nonanalytically when $\alpha$ approaches zero, and that the
avalanche-size distribution will not approach a power law with
increasing system size.
\end{abstract}


\maketitle

\section{Introduction}
\label{intro}
 The Olami--Feder--Christensen (OFC) earthquake model \cite{ofc92} is
probably the most studied nonconservative and supposedly
self-organized critical (SOC) model.  Systems are called self-organized
critical if they reach a stationary state characterized by power laws
without the need for fine-tuning an external parameter such as the
temperature. Many researchers in the field agree on confining the term
self-organized critical to those systems that are slowly driven and
that display fast, avalanche-like dissipation events. This means that
there is a separation of time scales, which can be interpreted as a
way of tuning a parameter to a small value \cite{man90}.

The prominent example for self-organized criticality is the sand-pile
model by Bak, Tang and Wiesenfeld \cite{bak87} (BTW), where it can be
shown analytically that the avalanche-size distribution is a power
law, implying a scale invariance: Avalanches of all sizes are due to
the same mechanism.  The BTW model satisfies a local conservation law,
which can naturally lead to power laws \cite{hwa89, gls90}, and
without local particle conservation the model is not critical
\cite{man90}. The mechanisms leading to SOC in nonconservative systems
are not yet well understood, and for the OFC model there is yet no
agreement on whether it is critical at all. While some authors find
critical behavior when going to larger systems sizes and employing
multiscaling methods \cite{lis01a,lis01}, others interpret similar
data as showing a breakdown of scaling \cite{mb03a}, and groups using
branching-ratio techniques claim to find what they call \emph{almost
criticality} \cite{car00, mb02}.

Despite the simplicity of its dynamical rules, the OFC model shows a
variety of interesting features that are unknown in equilibrium
physics and appear to be crucial for generating the apparent critical
(or almost critical) behavior.  Among these features are a marginal
synchronization of neighboring sites driven by the open boundary
conditions \cite{mid95}, and the violation of finite--size scaling
\cite{gras94,lis01} together with a qualitative difference between
system--wide earthquakes and smaller earthquakes \cite{lis01a}.  Also,
small changes in the model rules (such as replacing open boundary
conditions with periodic boundary conditions \cite{per96}, introducing
frozen noise in the local degree of dissipation \cite{mou96} or  in the threshold values \cite{jan93}, including lattice defects \cite{ceva95}),
destroy the SOC behavior. 
Recently, it was found that the results of computer simulations are
strongly affected by the computing precision \cite{dro02}, and that
the model exhibits sequences of foreshocks and aftershocks
\cite{her02, hhs04}. If energy input occurs in discrete steps instead
of continually and if thresholds are random but not quenched, one
finds quasiperiodicity combined with power laws \cite{ram06}.  The SOC
behavior fully breaks down in OFC systems in one dimension
\cite{dro05}, where only small and system-wide avalanches are
observed.

Since dynamics become extremely slow for large system sizes and for
 small values of the control parameter (implying strong dissipation)
 it is very difficult to obtain reliable results for the model based
 on computer simulations only. Thus, we find in the literature
 contradicting results concerning the transient time needed for the
 invasion of the `self-organized region` from the boundary into the
 middle of the system, and concerning the avalanche-size distribution.
 While the transient time is found by some authors to scale with
 system size with an
exponent depending on the level of dissipation \cite{mid95},
this exponent is found by others to be a
constant \cite{lis02}, while still others find that above some critical 
degree of dissipation
the invasion stops and never proceeds to the system's center \cite{gras94}.

Similarly, the avalanche-size distribution is found either to be a
power law with an universal exponent independent of the level of
dissipation for large enough system sizes (however, different values
for this exponent are reported in \cite{ceva98} and in
\cite{lis01,lis01a}), or a power law with a nonuniversal exponent
\cite{mid95,bot97}.  Some authors found no power law at all above a
critical degree of dissipation, but disagree on the value above which
no power laws occur \cite{ch00,cp01}. Still other authors suggest that
the dissipative OFC model is not critical at all (just like the
random-neighbor version of the model \cite{lj96}, which is a mean
field approximation \cite{ch97,bg97,ppk98}), but displays the new
feature of being close to criticality, as mentioned above.  If this is
correct, only the conservative case leads to power laws in the
distribution of avalanches \cite{car00, mb03a}.

By combining extensive computer simulations with analytical arguments,
we will in this paper propose a phenomenological theory for the
transient as well as the stationary behavior particularly in the
limit of large dissipation.  The most important conclusions are that
the transient time diverges nonanalytically when the control parameter
$\alpha$ approaches zero, and that the avalanche-size distribution
will not approach a power law with increasing system size.

The outline of the rest of this paper is as follows: In the next
section, we present the definition of the model and explain the
simulation algorithm.  Then, we investigate the transient dynamics
that brings the system from a random initial state to the stationary
state as function of the system size and the model parameter. Section
\ref{correlations} investigates the scaling behavior of the
self-organized patches displayed by the system in the stationary
state. The results flow into the interpretation of our simulation
results for the avalanche-size distribution, which is studied in
Section \ref{distribution}. Finally, in Section \ref{conclusion}, we
summarize and discuss our findings.
\section{The Model}
\label{model}
The OFC model originated by a simplification of the spring-block model
by Burridge and Knopoff \cite{bur67}.  To each site of a square
lattice we assign a continous variable $z_{ij}\in [0,1]$ that
represents the local energy.  Starting with a random initial
configuration taken from a constant distribution, the value $z$ of all
sites is increased at a uniform rate until a site $ij$ reaches the
threshold value $z_t=1$.  This site is then said to topple, which
means that the site is reset to zero and an energy $\alpha\times
z_{ij}$ is passed to every nearest neighbor.  If this causes a
neighbor to exceed the threshold, the neighbor topples also, and the
avalanche continues until all $z_{kl} < 1$. Then the uniform increase
resumes.  The number of topplings defines the size $s$ of an avalanche
or `earthquake`.  The coupling parameter $\alpha$ can take on values
in $(0,0.25)$. Smaller $\alpha$ means more dissipation, and
$\alpha=0.25$ corresponds to the conservative case.  Apart from the
system size $N$, the edge length of the square lattice, $\alpha$ is
the only parameter of the model. Except for  the initial condition, the
model is deterministic. After a transient time, the system reaches an
attractor of its dynamics. For periodic boundary conditions, the
attractor is marginally stable and has a period of $N^2$ topplings for
all $\alpha$ \cite{gras94,dro02}. All avalanches have the size 1, and
a site topples again only after all its nearest neighbors have
toppled. Measured in units of energy input per site, the period is
$1-4\alpha$. The behavior of the model is completely different for
open boundary conditions, where sites at the boundary receive energy
only from 3 or 2 neighbors and topple therefore on an average less
often than sites in the interior. This leads to the formation of
``patches'' of sites with a similar energy, and this patch formation
proceeds from the boundaries inwards. We are using open boundary
conditions throughout this paper.

Computer simulations of the model suffer from the long times needed to
reach the stationary state for large $N$ or small $\alpha$. Most of
the time is spent on searching for the site that will start the next
avalanche, i.e. for the site with the largest value of $z$.
Grassberger therefore used an algorithm that searches only among the
sites with the largest values of $z$ \cite{gras94}. In our
simulations, we used a different algorithm, based on a hierarchical
search. The system size is chosen to be a power of 2. The system is
divided into 4 boxes, each of which is again divided into 4 boxes,
etc., down to the box on the lowest level, which consists of 4 lattice
sites. Each box knows which of its 4 subboxes contains the site with
the largest $z$ value. Thus, the number of steps to find the site with
the largest $z$ value is $\log_2 N$, since 
after an avalanche only those
boxes have to be updated that have been affected by the avalanche.

\section{Transient Time}
\label{transient}

\begin{figure}
\includegraphics*[width=6.2cm]{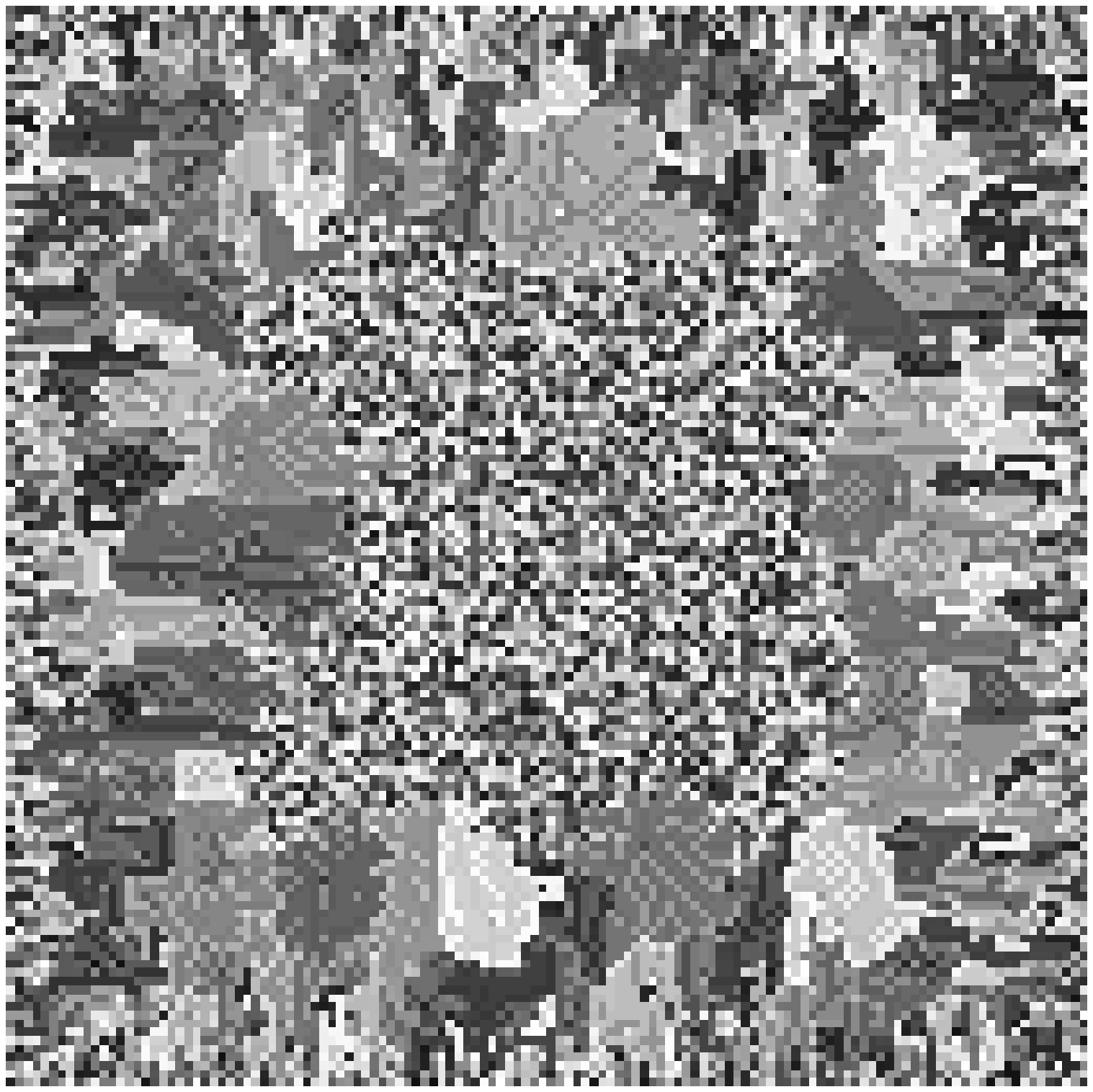}
\includegraphics*[width=6.2cm]{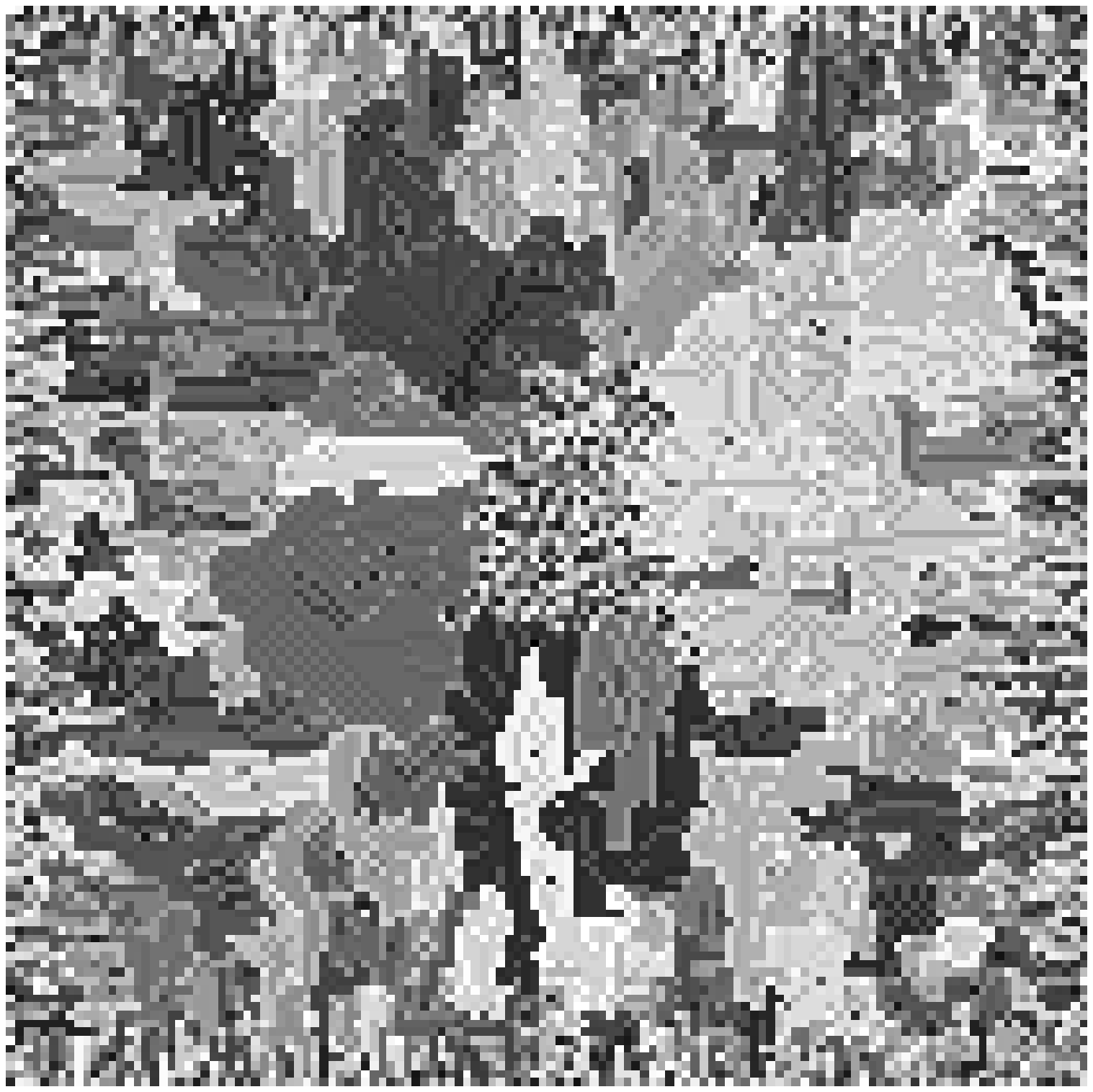}
\includegraphics*[width=6.2cm]{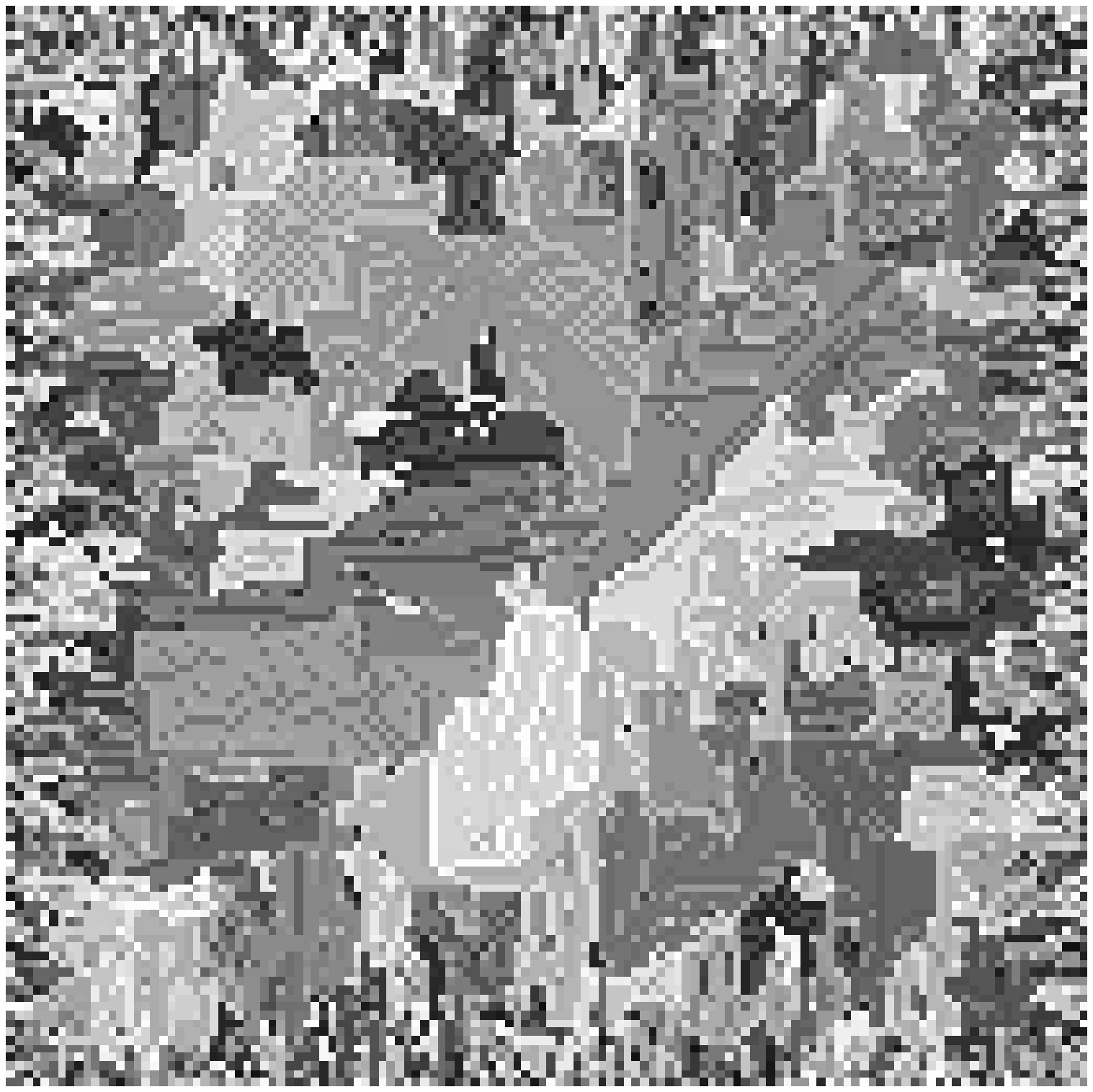}
\caption{Snapshots for a system with $N=128$, $\alpha=0.09$ after $10^5$, 
$5\times 10^5$  and $8\times 10^5$ topplings per site
\label{snapshot}}
\end{figure}

The transient time is the time needed for the patch formation to reach
the center of the system. Figure \ref{snapshot} shows 
a system with $N=128$ and $\alpha=0.09$ at three different times, the
last snapshot being taken in the stationary state.

One clearly distinguishes the patches close to the
boundaries and the disordered inner part of the system, which behaves
as if it was part of a periodic system. The time needed to establish a
patchy boundary starting from a random initial configuration is very
short, and virtually all of the transient time is needed to expand the
patchy region to the entire system. The patches become larger with
increasing distance from the boundary.

The first ones to investigate the transient behavior were Middleton
and Tang \cite{mid95}, who found that the transient time increases
as a power of $N$, with an exponent that depends on $\alpha$.
Later work on larger systems and for values of $\alpha$ larger than
0.15 by Lise \cite{lis02} found an exponent around $1.3$, 
which does not depend on $\alpha$.

\begin{figure}
\includegraphics[width=9cm]{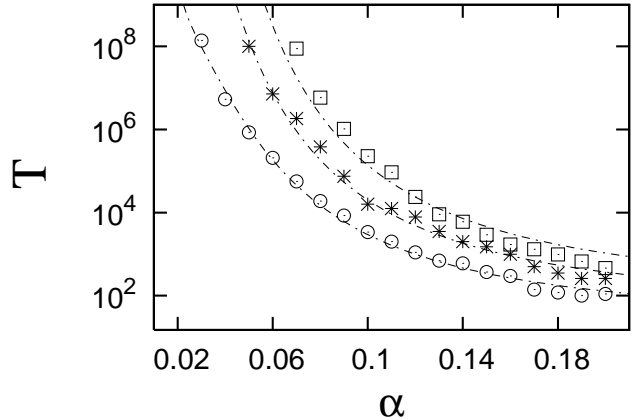}
\caption{
Time measured in topplings per site until the inner block vanishes for 
$N=64$ (circles), $N=128$ (stars) and $N=256$ (squares) as function of 
$\alpha$; the lines correspond to the function 
$T(\alpha, N) = \tilde{f}(\alpha)N^{\mu(\alpha)}$ as derived in the text.
\label{transient_tps}}
\end{figure}

We will argue that the exponent does indeed depend on $\alpha$, and
that it diverges for $\alpha \to 0$. Figure \ref{transient_tps} shows
our simulation results for the transient time for different system
sizes $N$ as function of $\alpha$.

Each data point is based on the simulation of one system. Averaging
over several initial conditions is not possible because of the long
computation times. The transient time increases with increasing
$\alpha$ and $L$, and it appears to diverge for $\alpha \to 0$. For
small $\alpha$, one might therefore obtain the impression that the
dynamics get completely stuck before the disordered block vanishes, as
was suggested by Grassberger \cite{gras94}. However, we found no solid
evidence and no good reason why this process should stop before the
patches fill the entire system. 
The dash-dotted lines in Figure \ref{transient_tps} are a fit with the 
Ansatz (\ref{generalpowerlaw2}), which is a generalized version of the result obtained in the following by using  mean-field like arguments.

We start with a local balance equation, which will lead us to an
expression for the toppling profile as function of time. This
consideration is similar to the one applied in \cite{dro05} to the
one-dimensional model. Let $t_{ij}$ be the mean number of topplings of
site $ij$ per unit time. 
If site $ij$ topples usually when $z_{ij}$ is
at the threshold (and not above), $t_{ij}$ must equal the mean amount
of energy that this site obtains per unit time. 
For small values of $\alpha$ this assumption is well satisfied. 
Let $g $ denote the rate of uniform energy input, and let $\tilde \alpha$ 
denote the average amount of energy passed to a neighbor during a toppling event. 
For small $\alpha$, the value of $\tilde \alpha$ deviates very little from
$\alpha$, and in the later part of this calculation we will therefore
replace $\tilde \alpha$ with $\alpha$. When discussing the size
distribution of avalanches further below, we will see that with
increasing system size the proportion of avalanches larger than 1
decreases towards zero, implying that the average amount of energy
passed to a neighbor approaches $\alpha$ even for larger values of
$\alpha$, and that most sites are exactly at the threshold when they
topple, as was already observered numerically in \cite{dro02, mb03}. 
The assumption that the value of $\tilde\alpha$ is constant
throughout time and throughout the system is a mean-field
assumption. Due to the approximations involved, we can 
expect that our theory makes predictions that are qualitatively
correct, but that the quantitative features could be different.

The balance equation  reads
\begin{equation}
t_{ij} = {g} + \tilde{\alpha} (t_{i+1,j}+t_{i-1,j}+t_{i,j-1}+t_{i,j+1})\, .
\end{equation}
Now, we have to take into account the structure of the system during
the transient time. The outer part consists of patches of different
sizes \cite{bot97}, and sites sitting in the same patch have to topple equally
often for the patch to persist for a long time (which is observed by
watching the system on the computer screen).
The value of $t_{ij}$
depends therefore on the distance to the boundary, which can be
measured in terms of the number of patches, $x$, between site $ij$ and
the boundary.  (We ignore here the fact that the system has corners,
which should not fundamentally change the argument. In any case, one
could consider a system that is periodic in one dimension and open in
the other, in order to avoid corners altogether.)  Sites in the
disordered block topple like in a system with periodic boundary
conditions, i.e. they receive the same input from all four
neighbors. For these sites we have therefore $t=g+4\alpha t$, or
\begin{equation}
\label{tnull}
t=t_0 \equiv\frac{g}{1-4\alpha}\, .
\end{equation}

In terms of the parameter $x$, the above balance equation for
the patchy part of the system becomes
\begin{equation}
t(x) = {g} + \tilde{\alpha} (t(x-1)+t(x+1)+ 2t(x))\, ,
\end{equation}
or, in a continuum notation
\begin{equation}
\frac{1-4\tilde{\alpha}}{\tilde{\alpha}}t(x)-\frac{d^2}{d x^2}t(x)-
\frac{{g}}{\tilde{\alpha}} = 0\, .
\end{equation}
The boundary conditions are $t(0)= 0$ ($x=0$ signifying the
non-existent neighbor of a boundary site) and $t(d) = t_0$,
with $d-1$ denoting the index of the patch next to the disordered block. 
The solution of the balance equation is then
\begin{equation}
\label{topplings}
t(x)=t_0\left(1-\frac{\sinh\left(\kappa (d-x)\right)}{\sinh\left(\kappa d\right)}\right)\, ,
\end{equation}
where $\kappa$ is given by $\kappa=\sqrt{(1-4{\alpha})/{\alpha}}$.

Next, we have to consider the advancement of the patchy structure into
the inner part of the system.  A site that is part of the inner block
can become part of a patch only if the difference of its energy value
$z$ to that of its outer neighbor is less than $\alpha$.  This
difference changes with time due to the different toppling rates.  The
patch next to the inner block topples less often than a neighbor of
that patch, which is part of the inner block, 
the difference in the
number of topplings per unit time being $t_0\left[\sinh(\kappa)
\right]/\left[\sinh\left(\kappa d\right)\right]$, which is
obtained from (\ref{topplings}) by inserting $x=d-1$.  
The difference in the number of topplings per unit time is identical to
the rate of change of the difference in the energy value $z$ between
the two neighbors. When this difference has increased by 1, it has
taken any intermediate value (in steps of size $\alpha$) and has
therefore certainly assumed a value smaller than $\alpha$. At that
moment, the site of the inner block becomes part of the patch.  The
time (or number of topplings per site) needed to add an additional
site to a patch is therefore proportional to
\begin{equation}
n_c(\alpha,d)  \sim \frac{\sinh\kappa d}{\sinh\kappa}\, .
\end{equation}
In the limit of small $\alpha$, $n_c(\alpha,d)$ is given by
\begin{equation}
\label{n_c}
n_c(\alpha,d)\sim \exp\left(\frac{d-1}{\sqrt{\alpha}}\right)\quad ,
\end{equation}
which has to be summed over all patches, 
weighted with the mean size of each generation of patches. 
The total transient time is therefore
\begin{equation}
T(\alpha,N) \sim \sum_{d=1}^{d_{max}(\alpha,N)} l(d) n_c(\alpha,d) 
\end{equation}
with $l(d)$ being the extension perpendicular to the boundary of a
patch of type $d$. Below in Section \ref{correlations}, we will see that 
$l(d) \sim Q(\alpha)^{d-1}$, where $Q$ is a function of $\alpha$ only and 
approaches 1 (from above) for $\alpha\to 0$. From the condition
\begin{equation}
\frac N 2 = \sum_{d=1}^{d_{max}} l(d) 
\end{equation}
we obtain then 
\begin{equation}
d_{max}(\alpha,N)\simeq\frac{\ln\left[\frac{N(Q-1)}{2q_0}+1\right]}{\ln Q}
\simeq \frac{\ln \frac{N(Q-1)}{2q_0}}{\ln Q}
\end{equation}
 for large  enough system sizes. $q_0$ is some constant 
(the extension of the patches of the first generation).
The result for small $\alpha$ is therefore 
\begin{equation}
\label{generalpowerlaw}
T(\alpha,N)\simeq\left(\frac{N(Q-1)}{2q_0}\right)^{\mu(\alpha)} \!
\exp\left(\frac{-2}{\sqrt{\alpha}}\right)
\end{equation}
with the exponent $\mu(\alpha)=1+\frac{1}{\sqrt{\alpha}\ln Q(\alpha)}\,$.
Using the ansatz $Q(\alpha)=\exp\left(f(\alpha)\right)$, also motivated in Section \ref{correlations}, 
with a leading term $f(\alpha)\simeq A\alpha^a$ and $A$ and $a$ positiv, yields
\begin{equation}
\label{mu_alpha}
\mu(\alpha)=1+\frac{1}{A\alpha^{a+0.5}}
\end{equation}
and
\begin{equation}
T(\alpha,N)\simeq\left(\frac{N}{2q_0}f(\alpha)\right)^{\mu(\alpha)}
\exp\left(\frac{-2}{\sqrt{\alpha}}\right) 
\end{equation}

Inspired by this result of the mean-field theory, we expect that 
the transient time is for small $\alpha$ given by an expression of the form
\begin{equation}
T(\alpha,N) \sim \tilde f(\alpha) N^{\mu(\alpha)}\,.
\label{generalpowerlaw2}
\end{equation}
The data shown in Figure \ref{transient_tps} agree with this expression.
The numerical values of the parameter are $A\sim 32.6$ and $a\sim 1.262$.
$\tilde f(\alpha)$ was fitted in the form
$\tilde f(\alpha)\sim\exp\left[-V(\alpha+0.01)^v+D\right]$, but any other 
expression could be equally valid. 
We would like to stress that this Ansatz was motivated by the mean field 
exponent $\mu$ for the leading dependence on $N$, which we think mirrors the 
true behavior correctly.

\section{Correlation function and correlation length}
\label{correlations}

As has become clear from the previous section, the size distribution
of patches as function of $N$ and $\alpha$ and of their distance from
the boundary is an important feature of the system. It affects not
only the transient time, but also the avalanche size distribution,
which will be discussed in the next section.

We therefore investigate in this Section how the extension of the
patches in the directions parallel and perpendicular to the boundary
increases with the distance from the boundary. For this purpose, we
evaluate the correlation function
\begin{equation}
C(r) = \langle (z_{ij} - z_{i,j+r})^2\rangle-\langle z_{ij}\rangle^2
\end{equation}
for a fixed distance $i$ from the boundary for different times,
starting again at a random initial configuration. We performed our
simulations with systems that are periodic in the direction of the
second coordinate, i.e. site $j+N$ is identical to site $j$. We chose
$N=2^{15}$ in order to obtain good statistics. The length of the
system in the other direction was chosen just as large as needed,
between 48 (for small $\alpha$, where the invasion front proceeds very
slowly) and 512 (for large $\alpha$).

\begin{figure}[ht]
\includegraphics[width=9cm]{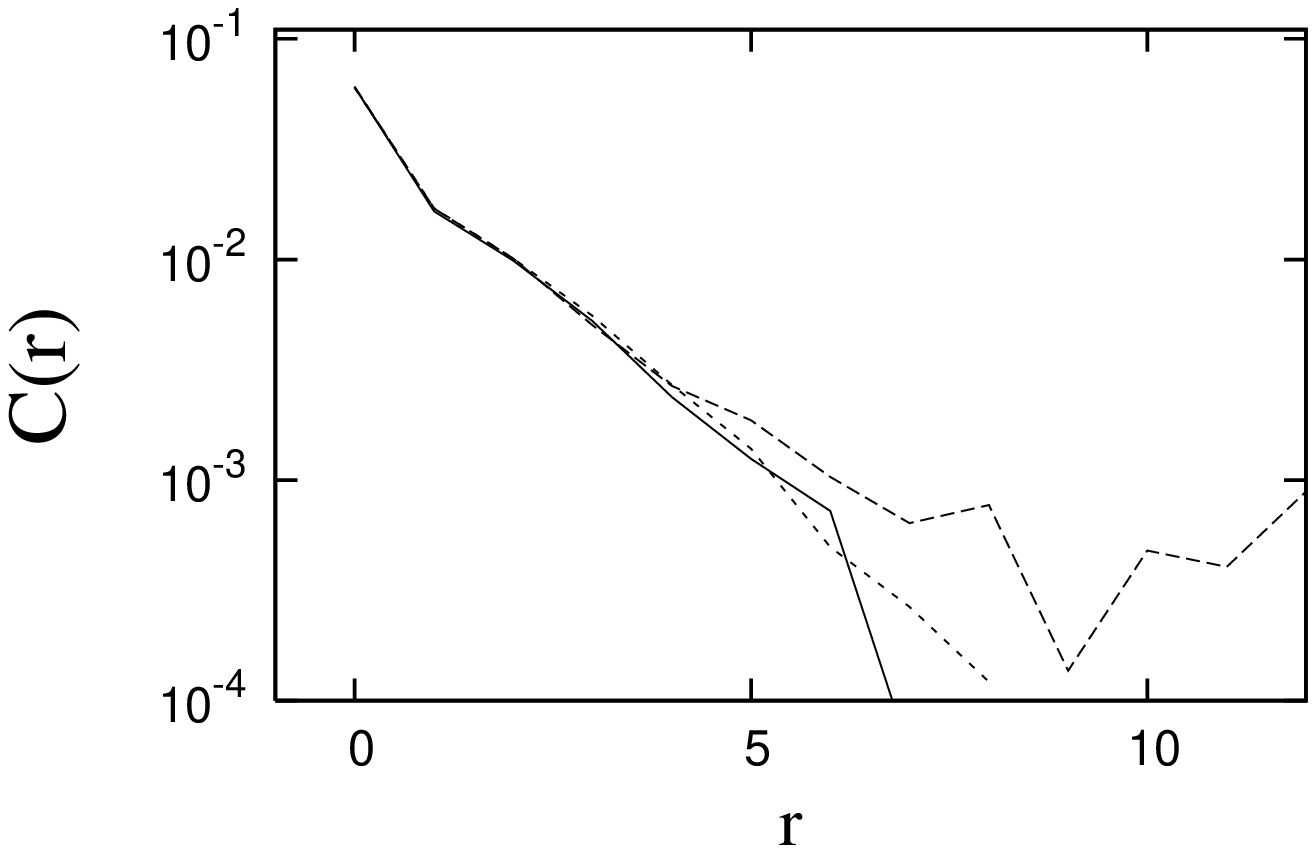}
\includegraphics[width=9cm]{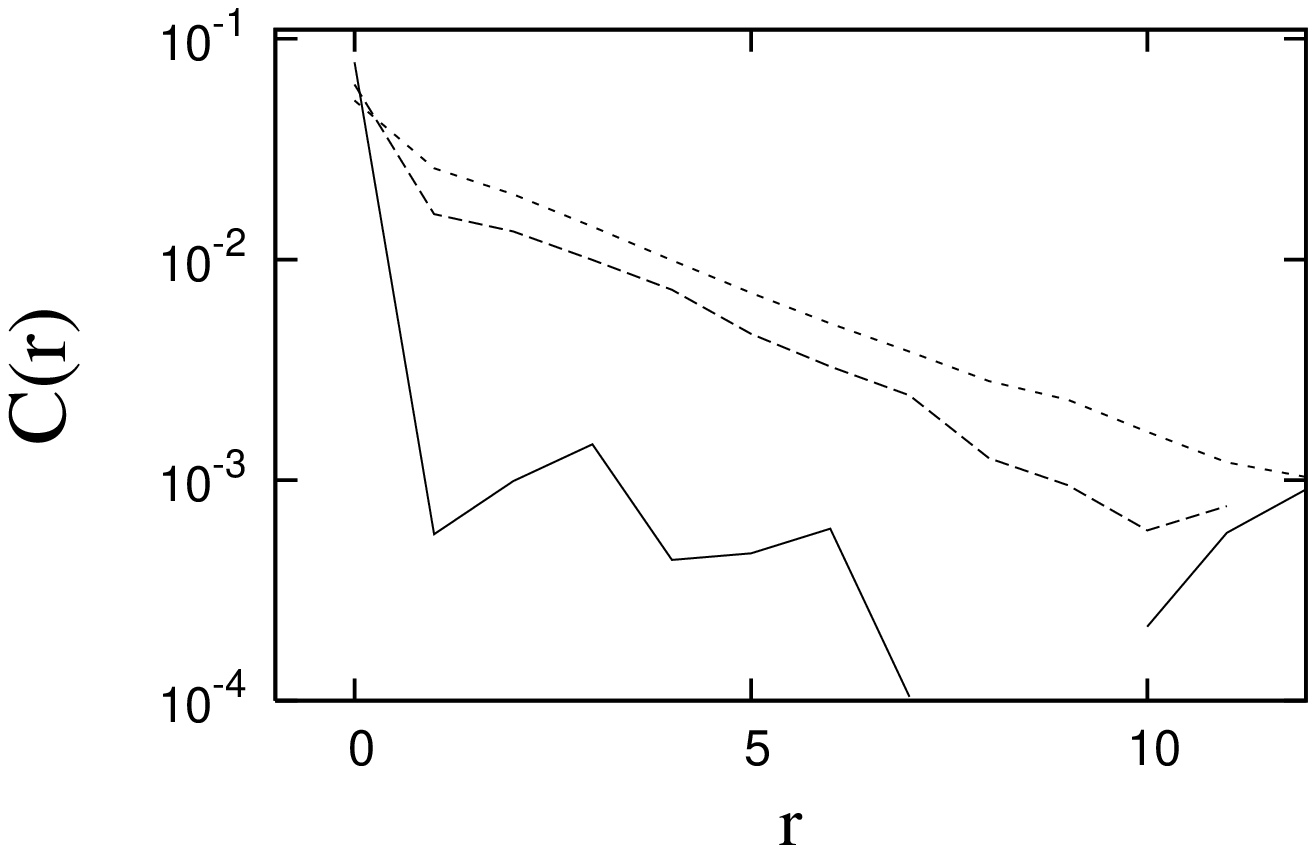}
\caption{
Correlation function $C(r)$ for $\alpha=0.08$ for a distance $d=10$ sites 
(top) and $d=20$ sites (bottom) from the boundary at three different times 
(after $2570$ (solid line), $5130$ (dashed line) and  $12490$ (dotted line)
topplings per site). 
\label{cor_function}
}
\end{figure}

\begin{figure}[ht]
\includegraphics[width=9cm]{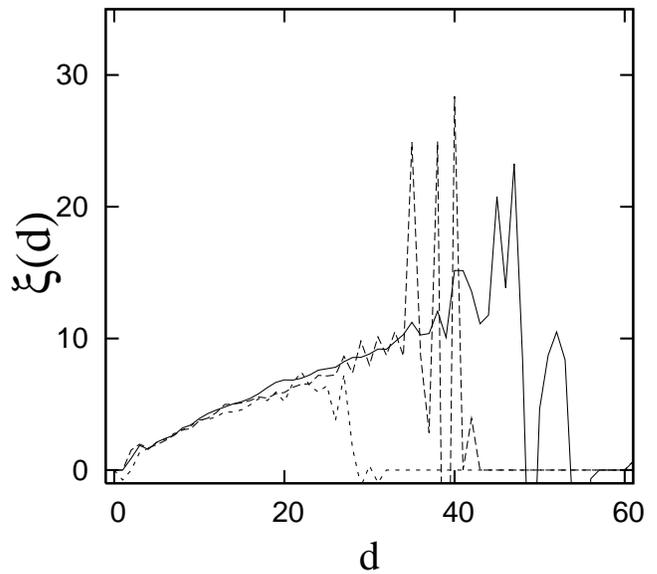}
\caption{
Correlation length $\xi$ as function of the distance $d$ to the boundary for $\alpha=0.12$ after 650, 1290 and 2570
topplings per site.
\label{cor_lenght}
}
\end{figure}

Figure \ref{cor_function} shows the correlation function for
$\alpha=0.08$ at distance 10 and distance 20 (measured in number of sites)
from the boundary for three different times. 
One can see that at distance 10 the correlation function does not change 
any more with time, which means that the patch structure has been 
established at least up to this depth before the first measurement. 
We can furthermore conclude that the typical scale
of the patches at a given distance from the boundary does not change
any more when new patches are formed further inside. 
At distance 20, we see that the correlation function builds up with
time from zero to an exponentially decaying function $C(r) \sim
e^{-r/\xi}$.  
Figure \ref{cor_lenght} shows the correlation length
$\xi$ for $\alpha=0.12$ as function of the distance to the boundary
for three different times. In the region where the patches are already
present, $\xi$ increases as a power law in the distance from the
boundary, and then falls down to zero. We see again that $\xi$
remains constant once the patches have emerged.
The large fluctuations seen before the decrease to zero occur in the 
region where the patches are just being formed. Due to large fluctuations 
in space, the averaging over the length of the system does not lead to a
smooth curve for the system sizes used. 

\begin{figure}[ht]
\includegraphics[width=9cm]{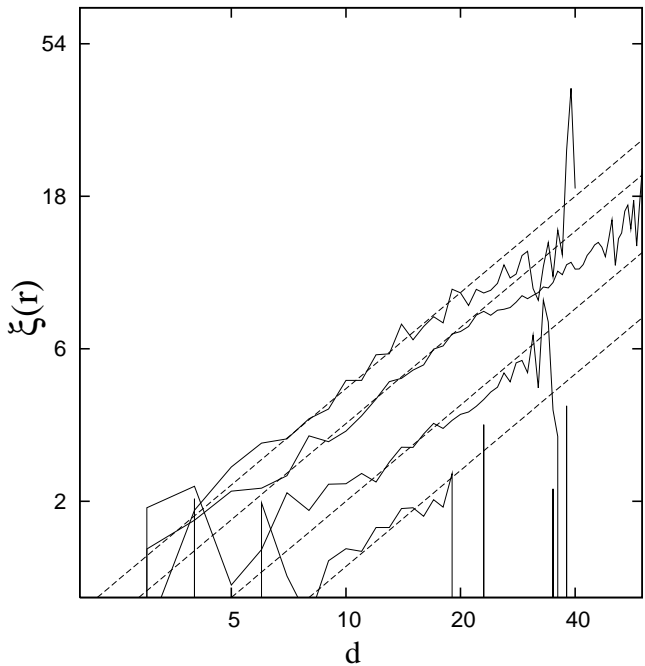}
\caption{
Correlation length $\xi$ as function of the distance to the boundary
for $\alpha=0.06, 0.09, 0.12$ and $0.15$ (from bottom to top) at the
largest times simulated for a given value of $\alpha$ (solid lines); 
The dashed lines correspond to lines of slope 1
\label{cor_lenght_a}
}
\end{figure}

Figure \ref{cor_lenght_a} shows the correlation length as function of
depth for different values of $\alpha$.
The data are in good agreement with a linear increase of $\xi$ with
the distance $d$ from the boundary, but with a factor that decreases with
decreasing $\alpha$. However, we cannot rule out a power law 
$\xi \sim d^\eta$ with an exponent $\eta <  1$ that increases with $\alpha$. 

\begin{figure}[ht]
\includegraphics[width=6cm]{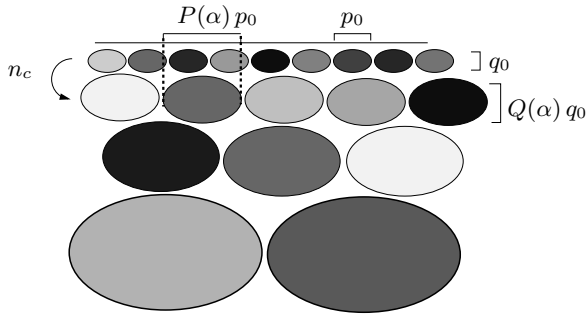}
\begin{picture}(0,0)
\put(-190,90){$n_c$}
\put(-126,110){$P(\alpha)\, p_0$}
\put(-64,110){$p_0$}
\put(-8,93){$q_0$}
\put(-1,75){$Q(\alpha)\,q_0$}
\end{picture}
\caption{
Schematic view of the system's structure: the width and height of
patches increase with a power law in the distance to the boundary. 
Different generations of patches are coupled via $n_c$, the increase
in the size of the patches is $P(\alpha)$ parallel to the boundary
and $Q(\alpha)$ perpendicular to it, starting with a size $s_0=p_0q_0$.
\label{setup}
}
\end{figure}
The linear (or power-law) increase of the correlation length together with the
patchy structure leads to the following schematic picture (see Fig.~\ref{setup}):

The characteristic size of patches increases with distance from the
boundary. From one generation of patches to the next,
the width and height of the patches increase with factors $P(\alpha)$
and $Q(\alpha)$ respectively. In the case $\eta=1$, we have $P=Q$. (Of
course, the patches at a given distance from the boundary do not all
have exactly the same size, but a size of the indicated order of
magnitude.)  From snapshots of the systems, it is clear that
$P(\alpha)$ and $Q(\alpha)$ increase with $\alpha$. 
Furthermore, there must be a lower bound of 1 to both factors in the 
limit $\alpha\to 0$. 
Thus, we can write $Q(\alpha)=\exp(f(\alpha))$ with a monotonically increasing 
function $f(\alpha)$ and $f(0)=0$. The leading dependence on $\alpha$ can be expected to be $f(\alpha)=A\alpha^a$ with positive $A$ and $a$.

The correlation length in the $i$th generation of patches (counting from
the boundary) is
\begin{equation}
\xi \sim P^i \sim d^\eta
\end{equation}
and the distance from the boundary is
\begin{equation}
d\sim \sum_{j=1}^iQ^j\, .
\end{equation}

Based on this picture, we can write down an expression for the size
distribution of patches, which will be an important tool when
discussing the size distribution of avalanches. A line at the distance
$d$ from the boundary cuts through $\sim N/\xi h$ new patches of width
$\xi$ and height $h\sim \xi^{1/\eta}$ through which the neighboring
line does not cut, since there are $\sim N$ sites along this line. The
width distribution of patches is therefore given by
$$n_P(\xi) d\xi = \frac{N}{\xi h} d \, (d)\, ,$$
leading to
\begin{equation}
\label{nPxi}
n_P(\xi) \sim \frac{N}{\xi^2}\, ,
\end{equation}

Transforming this into the size distribution $n_P(s)$ with $s \sim \xi h$, 
we obtain
\begin{equation}
n_P(s) \sim N s^{-\frac{1+2\eta}{1+\eta}}\, .\label{nPmu}\end{equation} In the
likely case that $\eta=1$, we have an exponent $-3/2$ in the size
distribution of patches. Expressed in terms of $P$ and $Q$ instead of $\eta$, the last equation becomes
\begin{equation}
n_P(s) \sim N s^{-\frac{\ln P}{\ln PQ}-1}\, ,\label{nPPQ}\end{equation} where $P$
and $Q$ depend on $\alpha$. This expression can also be obtained
directly from the recursion relation
\begin{equation}
\int n_P\left(sP(\alpha)Q(\alpha)\right) ds = \int n_P(s)
\frac{1}{P(\alpha)}ds\, ,\label{nPstart}
\end{equation}
where the integral is taken over one generation of patch sizes.

\section{Avalanche size distribution}
\label{distribution}
Now we turn to the size distribution of avalanches in the stationary
state. We made sure that the process of patch formation has reached
the center of the system, before we evaluated the avalanche size
distribution. 

In view of the results in Section \ref{transient}, 
we are now in the position to check how trustworthy
the results reported in the literature are.
As was already pointed out by Grassberger \cite{gras94}, 
transient times are extremely long, and the first publications \cite{ofc92,cob92, oc92, co92} can have considered stationary systems only for the largest values of $\alpha$.

It appears that many avalanche size distributions presented in the
last decade were actually obtained during the transient stage. We can
check this only when the authors state how many initial avalanches
they discarded for given $N$ and $\alpha$. Unfortunately, not all
authors write how they decided if the system is in the stationary
state.  By observing statistical properties and comparing them at
different times, one can be mislead to believe that the system has
become stationary, although the advancement of the patches has only
become very slow. Generally, the larger $\alpha$ and the smaller the
system size, the more likely is it that the published avalanche size
distributions were obtained in the stationary state. For example, the
results published in \cite{ceva95,ceva98} with $L=25,45$ were probably
not taken in the stationary state for $\alpha$ below $0.2$.  Even
Grassberger was evaluating avalanche size distributions during the
transient stage in some parameter regimes.

While taking small system sizes has the advantage of reaching the
stationary state fast, they have the disadvantage of being strongly
affected by finite-size effects. It is therefore very difficult to
predict the avalanche-size distributions in the thermodynamic limit. 

\begin{figure}[ht]
\includegraphics[width=9cm]{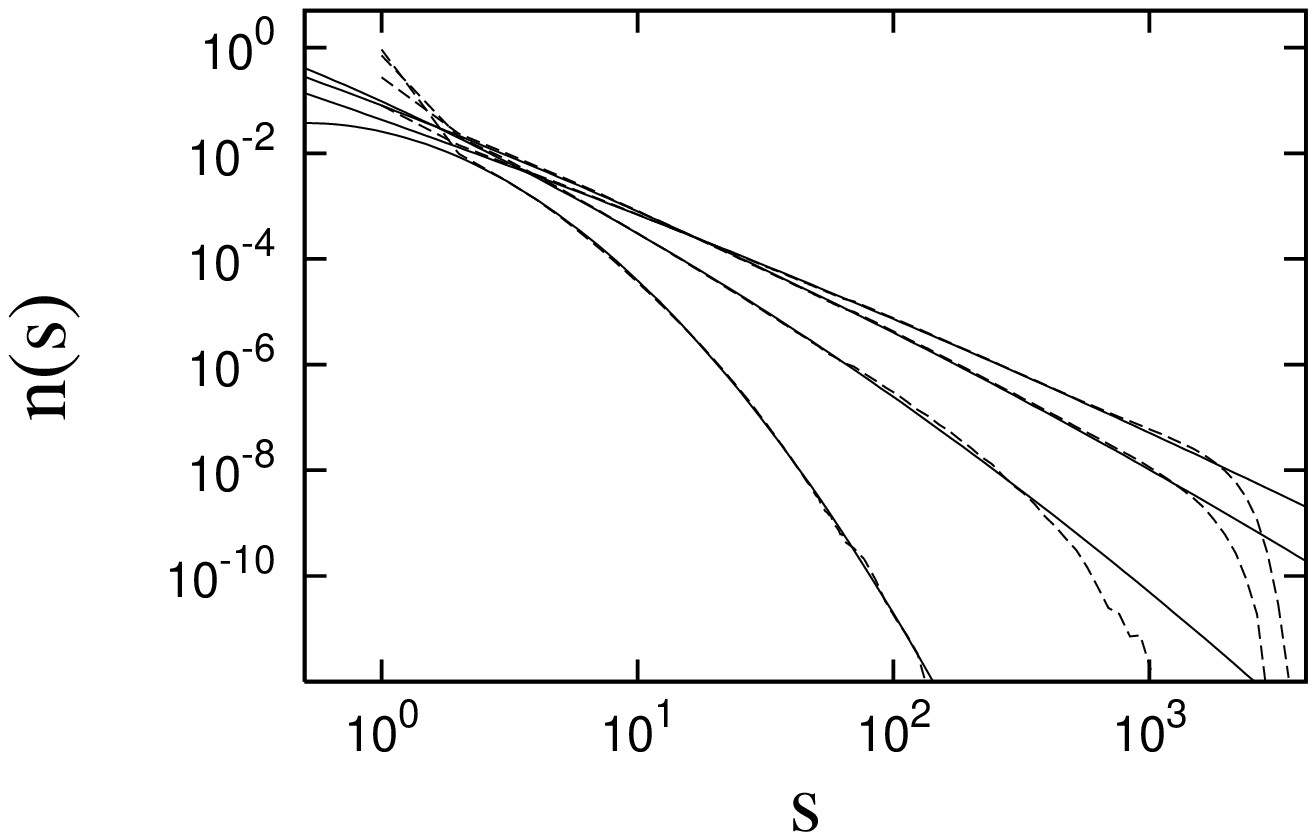}
\includegraphics[width=9cm]{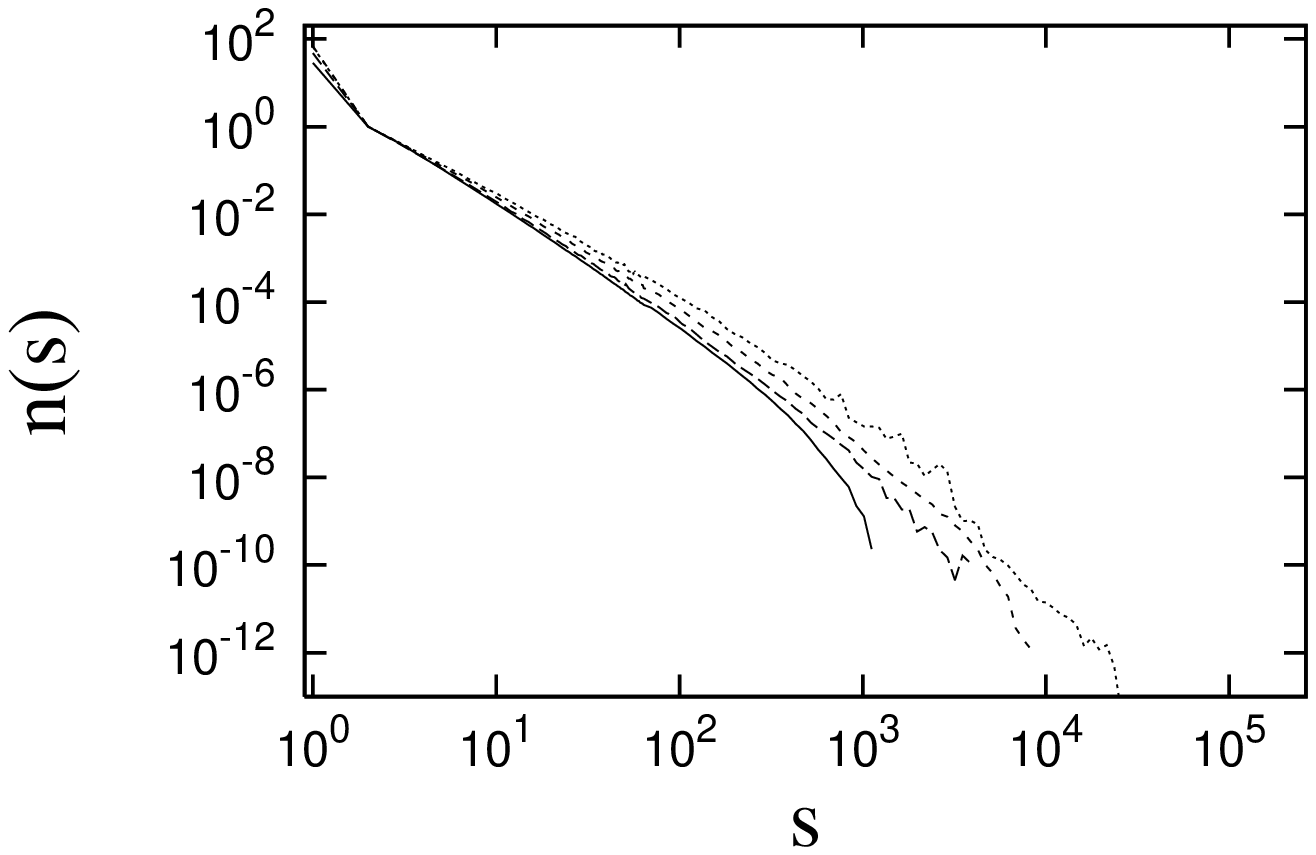}
\includegraphics[width=9cm]{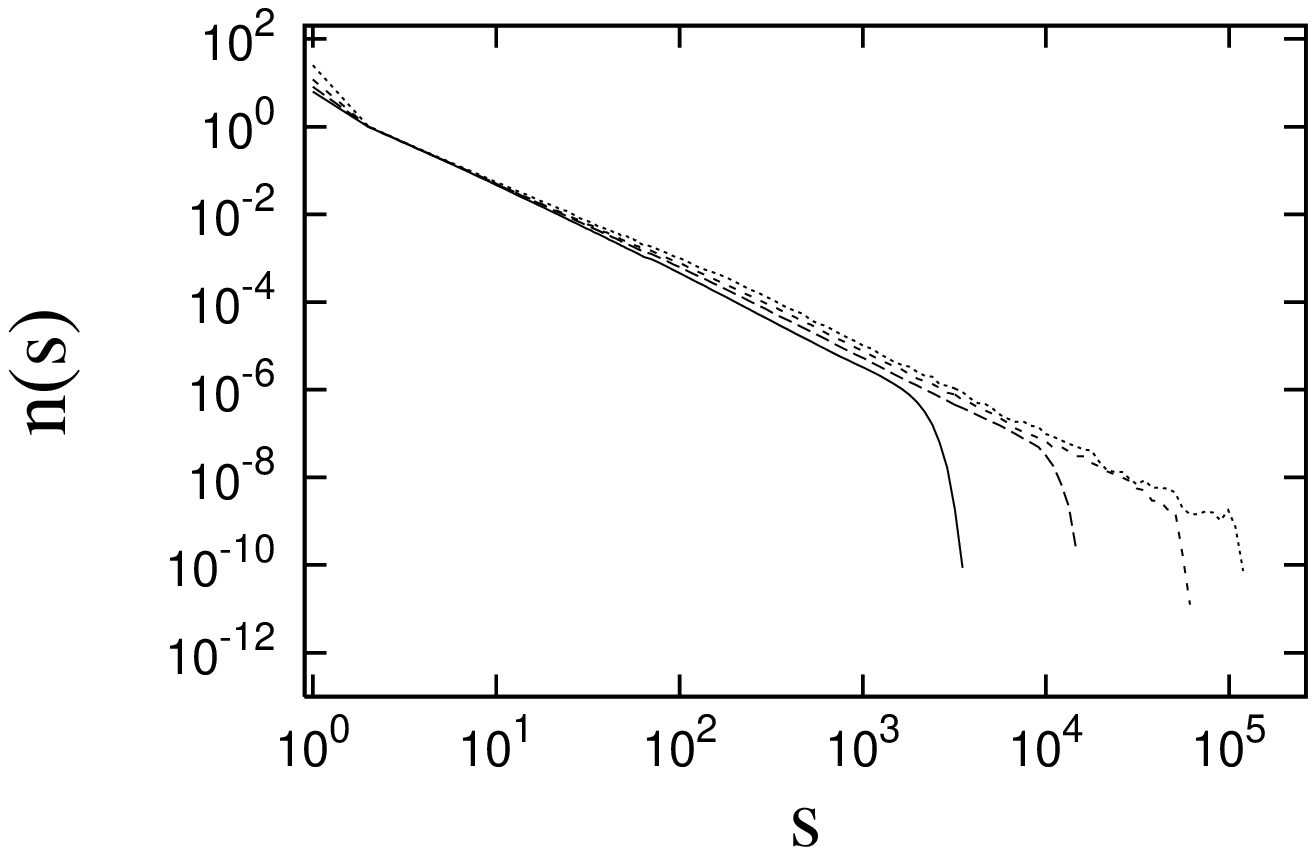}
\caption{
Size distribution of avalanches for different parameter;
top: system size $N=64$ and $\alpha=0.03,0.08,0.13,0.18$, steeper curves correspond to smaller $\alpha$;
the $s$-axxis extends up to the total number of sites $4096$; 
the distributions are normed on the total number of topplings;
solid lines correspond to $f(s)\sim s^{-\tau -\sigma \ln s}$. 
middle and bottom: size distribution for system sizes $N=64, 128, 256, 512$; 
$\alpha=0.09$ (middle) and $\alpha=0.17$ (bottom); 
the distributions are divided by $n(2)$.
\label{sd}
}
\end{figure}

Figure \ref{sd}  shows avalanche size distributions for varying $\alpha$
with fixed $N$ and for varying $N$ with fixed $\alpha$. The value of
$N$ in the top figure has been chosen small enough that the system
could reach the stationary state even for the smallest value of
$\alpha$, which was 0.03. We can discern the following features:
\begin{enumerate}
\item At least for value of $\alpha$ smaller than 0.17, the avalanche size distribution is no power law. A fit of the form $n(s)\sim s^{-\tau(\alpha) -\sigma(\alpha)\ln s}$ approximates the data much better than a pure power law. 
\item $n(s)$ changes its shape with increasing $N$, implying that the system size affects the relative weight even of small avalanches, at least fo the system sizes considered. This effect is stronger for smaller $\alpha$. Only for the largest value of $\alpha$ is the main effect of the finite system size a rather sharp cutoff at $N^2$.
\item The weight of avalanches of size 1 increases with increasing system size, while the weight of all larger avalanches decreases as $1/N$ (see below).
\end{enumerate}

In the following, we will explain these features based on the results
obtained in the previous sections, and on what is known from
literature.  Described in words, the scenario is the following:
Patches persist for a long time before they change their shape \cite{bot97},
due to an avalanche that enters the patch from outside \cite{gras94},
and patches further inside the system are rearranged less often.
Large, patch-wide avalanches are mainly triggered at the boundaries of
the system. Whenever a patch-wide avalanche took place, there is a
sequence of `aftershocks` with decreasing size according to Omori's
law \cite{her02}, and after a short time there occur mostly single
topplings within a patch, until the next large avalanche comes from a
patch of the previous generation.

Let us quantify these statements. Analogous to the process of
synchronizing neighboring sites discussed in Section \ref{transient},
neighboring patches also need a certain number $n_c(\alpha)$ of
patch-wide avalanches in the patch closer to the boundary, before the
inner patch experiences a patch-wide avalanche. This can be evaluated
using Eq.~(\ref{topplings}) for the situation that $d = d_{max}$.
We therefore obtain the recursion relation (compare (\ref{nPstart}))
\begin{equation}
\int n_{pw}\left(P(\alpha)Q(\alpha)s\right) ds = \int n_{pw}(s)
\frac{1}{P(\alpha)n_c(\alpha)}ds\, ,
\label{nsstart}
\end{equation}
for the size distribution of patch-wide avalanches. 
If $n_c$ was independent of the generation number $i$, 
this would result in a power law
 $n_{pw}(s)\propto N s^{-\tau(\alpha)-1}$
with an exponent
\begin{equation}
\label{tau}
\tau(\alpha)=\frac{\ln P(\alpha)n_c(\alpha)}{\ln P(\alpha)Q(\alpha)}\quad.
\end{equation}

For systems not too big, and for large enough $\alpha$, 
there are only a few generations, and the approximation of a constant $n_c$ is not too bad. 
Evaluating  Eq.~(\ref{topplings}) for small $\alpha$, we obtain 
the following result for $n_c$ that depends on the generation index $i$, 
\begin{equation}
\label{nci}
n_c(\alpha,i)\sim \exp(\frac{i-1}{\sqrt{\alpha}})\, ,
\end{equation}
(see also equation (\ref{n_c})).
Iterating Equation (\ref{nsstart}), we need to evaluate the 
product 
\begin{equation}
\label{prod}
\prod_{j=1}^i \left(\frac{1}{P(\alpha)n_c(\alpha,j)}\right)=\left(\frac{1}{P(\alpha)}\right)^i \exp{\left(-\frac{i(i-1)}{2\sqrt{\alpha}}\right)}\quad,
\end{equation}
which leads (using $\ln i \sim \ln s / \ln PQ$) to size distribution
of patch-wide avalanches of the form
\begin{equation}
n_{pw}(s)\sim Ns^{-\tau(\alpha) -1 -\sigma(\alpha)\ln s}\quad,
\end{equation}
where $\tau(\alpha)$ and $\sigma(\alpha)$ are given by
\begin{eqnarray}
\label{exponents}
\tau(\alpha)&=&\frac{1}{\ln (P(\alpha)Q(\alpha))}\left(\ln P(\alpha)-\frac{1}{2\sqrt{\alpha}}\right)\nonumber\\
\sigma(\alpha)&=&\frac{1}{2\sqrt{\alpha}(\ln(P(\alpha)Q(\alpha)))^2}\label{sigmatau}
\end{eqnarray}

\begin{figure}[ht]
\includegraphics[width=9cm]{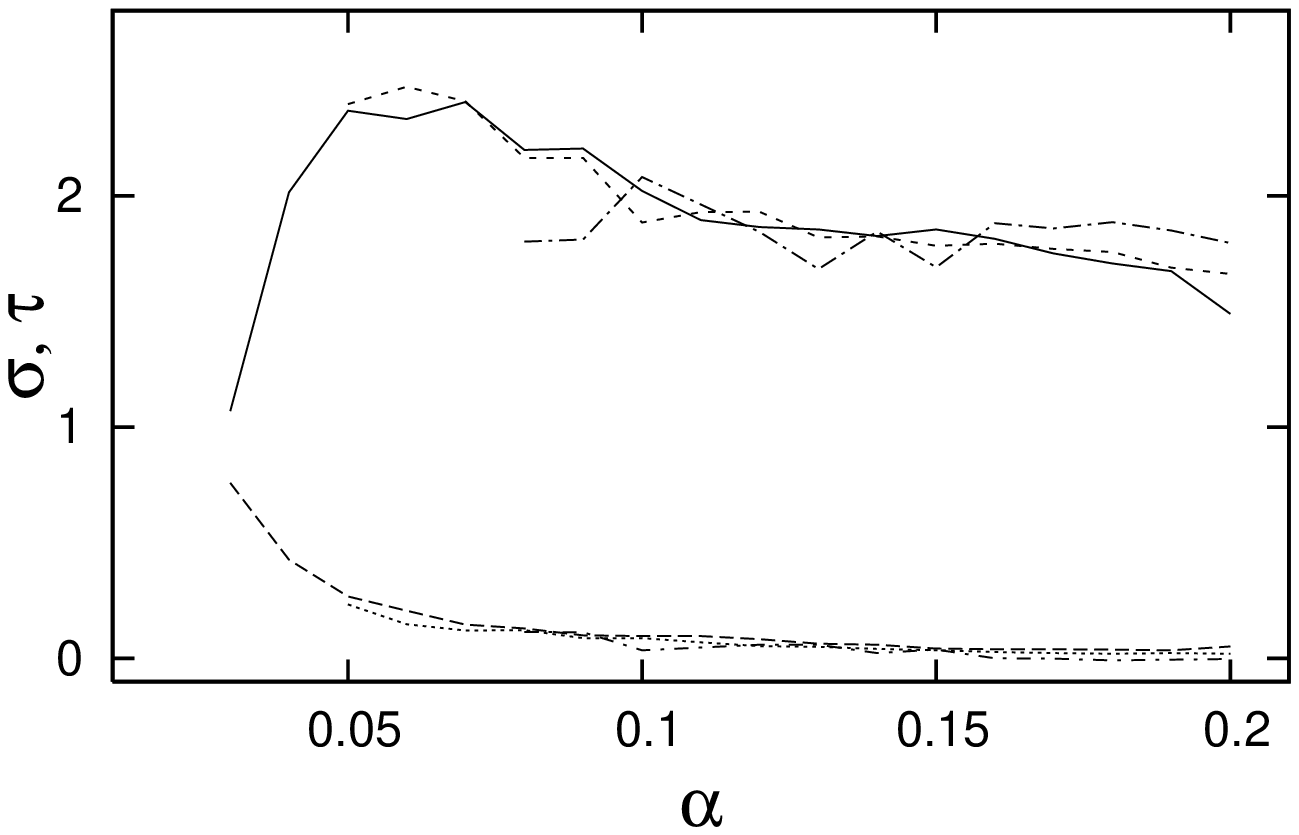}
\caption{
The coefficients $\sigma(\alpha)$ (lower set of curves) and $\tau(\alpha)$ (upper set) as function of $\alpha$ as found by fitting the distributions $n(s)$ for $N=64, 128$ and $256$for those values of $\alpha$ where stationary systems were reached.
\label{exp_fit}
}
\end{figure}

Now, we have to estimate the effect of `aftershocks` on the size
distribution of avalanches. These aftershocks will lead to an
avalanche-size distribution that differs from that of the patch-wide
avalanches. Aftershocks are avalanches that occur within a patch after
a patch-wide avalanche. We assume that their size distribution is a power law with a cutoff at the size of the patch. This is motivated by the finding that systems that are dominated by one large patch display a power-law size distribution of avalanches (see also \cite{her02, hhs04}). 
Therefore, we set
$$n_{as}(s | s') = s'^{\tau_0} s^{-\tau_0} \theta(s'-s)\, ,$$ with
$n_{as}(s | s')$ being the number of aftershock avalanches of size $s$
in a patch of size $s'$, and with an exponent $\tau_0$, which has a
value around $1.8$ (i.e.,  the value found in \cite{lis01} for
systems that have essentially one large patch).  The size distribution
of avalanches is then given by
\begin{eqnarray}
n(s) &\propto& N\int_s^\infty n_{pw}(s')n_{as}(s|s') ds'\nonumber\\
&=& Ns^{-\tau_0}\int_s^\infty  s'^{-\tau(\alpha) -1 -\sigma(\alpha)\ln
  s'}s'^{\tau_0}\nonumber\\
&\sim& Ns^{-\tau(\alpha) -\sigma(\alpha)\ln  s}\label{nsend}
\end{eqnarray}
apart from a factor containing terms that depend on  $\ln s$.
Thus, the avalanche-size distribution is not a power law, but it has
an exponent that depends logarithmically on $s$. As we have shown
above, the data agree well with such a law. Figure \ref{exp_fit} shows
our results obtained for the coefficients $\sigma$ and $\tau$ by
fitting the avalanche-size distribution with the expression
(\ref{nsend}). Although we can expect that the data are affected by
finite-size effects particularly for small $\alpha$, we see that the
functions $\sigma(\alpha)$ and $\tau(\alpha)$ show a behavior that is
in agreement with our expressions (\ref{sigmatau}): For small
$\alpha$, $\tau$ decreases rapidly and will eventually become
negative, while $\sigma$ tends to large positive values.

The cutoff of the avalanche-size distribution is determined by the
size of the largest patch. As this size becomes smaller with smaller
$\alpha$, the cutoff descreases also. Furthermore, since larger
patches make a contribution to smaller avalanches via aftershocks, the
effect of the finite system size will be felt down to avalanche sizes
much smaller than the largest patch. This is what is observed in the
data. 

Finally, let us discuss the weight of avalanches of size 1. After a
patch-wide avalanche and the resulting aftershocks, a patch has single
topplings (i.e., avalanches of size 1), just as a system with periodic
boundary conditions, until a new patch-wide avalache comes from
outside. The total number of avalanches of size larger than 1 per unit
time is given by
$$\int_2^\infty n(s) ds \propto N\, .$$
However, the total number of topplings per unit time is proportional
to the number of sites in the system $N^2$. We conclude, that only a
proportion of the order $1/N$ of avalanches has a size larger than 1. 

\section{Conclusion}
\label{conclusion}

In this paper, we have shown that the transient time for the OFC model
increases as function of the system size $N$ and the coupling
parameter $\alpha$ as $T(\alpha, N)\sim N^{\alpha^{-\mu}}$, 
apart from corrections depending on $\alpha$ which do not affect this
leading non-analytical behavior. This finding is in contrast to
earlier predictions that the trasient time increases as a power law
with system size, or that the transient time becomes infinite when
$\alpha$ is smaller than some value. We obtained these results by
performing a mean-field like calculation for the number of topplings per
site and for the advancement of the patchy structure into the inner
part of the system. 

Furthermore, by evaluating the correlation length of the energy values
we found that the size of the patches increases as a power law with
the distance from the boundary, leading to power law size distribution
of the patches. Even if we assume that the size distribution of
avalanches within a patch is a power law, we find based on the results
mentioned so far that the overall size distribution of avalanches is
no power law, but has a logarithmic dependence in the exponent on the
avalanche size, i.e., is of a log-Poisson form. This finding is
supported by the simulation data and is valid at least for smaller
$\alpha$, where the system is not dominated by one large patch.

We obtained our simulation results by using an efficient algorithm,
however, the sharp increase of the transient time with system size
especially for small $\alpha$ made it impossible to study system sizes
as large as necessary to see the true asymptotic behavior of the
avalanche size distributions. 

Our findings are interesting for several reasons. First, the OFC model
appears to show many features found in real earthquakes. As far as
earthquake predictability \cite{pc94} or Omori's law \cite{her02, hhs04}
are concerned, this model appears to be closer to reality than others.
If $\alpha$ is chosen above $0.17$, the avalanche size distribution agrees 
best with the Gutenberg-Richter law \cite{lis01}. 
Second, the OFC model demonstrates that
apparent power laws need not reflect a true scale invariance of the
system. We expect that this is true for many natural driven
systems. Due to the dynamics of the model, there occur avalanches of
all sizes, however the mechanisms producing these avalanches are
different on different scales. Large avalanches are mainly patch-wide
avalanches, while smaller avalanches occur within patches during a
series of foreshocks or aftershocks. Also, avalanches at different
distance from the boundaries have different sizes. The observed
``power laws'' are thus \emph{dirty} power laws, which appear like
power laws over a wide range of parameters and over a few decades on
the avalanche size axis, while the ``true'' analytical form is no
power law. Third, the lack of a true scale invariance is accompanied
by a decreasing weight of avalanches larger than 1 with increasing
system size. This indicates again that the avalanche size distribution
of the model does not approach some asymptotic shape with increasing
system size, but that the weights of different types of avalanches
shift with the system size. This effect has most clearly been seen in
one dimension, where the distributions split into a $\alpha$
dependent part at small avalanche sizes and a peak at sizes of order
of the system size.  Fourth, the extremely long transient times point
to the possibility the some driven natural systems with avalanche-like
dynamics are not in the stationary regime either.

\begin{acknowledgments}
This work was supported by the Deutsche 
Forschungsgemeinschaft (DFG) under Contract No Dr300/3-1 and Dr300/3-2.
\end{acknowledgments}

\bibliography{mybib}

\begin{thebibliography}{36}
\expandafter\ifx\csname natexlab\endcsname\relax\def\natexlab#1{#1}\fi
\expandafter\ifx\csname bibnamefont\endcsname\relax
  \def\bibnamefont#1{#1}\fi
\expandafter\ifx\csname bibfnamefont\endcsname\relax
  \def\bibfnamefont#1{#1}\fi
\expandafter\ifx\csname citenamefont\endcsname\relax
  \def\citenamefont#1{#1}\fi
\expandafter\ifx\csname url\endcsname\relax
  \def\url#1{\texttt{#1}}\fi
\expandafter\ifx\csname urlprefix\endcsname\relax\def\urlprefix{URL }\fi
\providecommand{\bibinfo}[2]{#2}
\providecommand{\eprint}[2][]{\url{#2}}

\bibitem[{\citenamefont{Olami et~al.}(1992)\citenamefont{Olami, Feder, and
  Christensen}}]{ofc92}
\bibinfo{author}{\bibfnamefont{Z.}~\bibnamefont{Olami}},
  \bibinfo{author}{\bibfnamefont{H.~J.~S.} \bibnamefont{Feder}},
  \bibnamefont{and}
  \bibinfo{author}{\bibfnamefont{K.}~\bibnamefont{Christensen}},
  \bibinfo{journal}{Phys.\ Rev.\ Lett.} \textbf{\bibinfo{volume}{68}},
  \bibinfo{pages}{1244} (\bibinfo{year}{1992}).

\bibitem[{\citenamefont{Manna et~al.}(1990)\citenamefont{Manna, Kiss, and
  Kert\'esz}}]{man90}
\bibinfo{author}{\bibfnamefont{S.}~\bibnamefont{Manna}},
  \bibinfo{author}{\bibfnamefont{L.}~\bibnamefont{Kiss}}, \bibnamefont{and}
  \bibinfo{author}{\bibfnamefont{J.}~\bibnamefont{Kert\'esz}},
  \bibinfo{journal}{J.\ Stat.\ Phys.} \textbf{\bibinfo{volume}{61}},
  \bibinfo{pages}{923} (\bibinfo{year}{1990}).

\bibitem[{\citenamefont{Bak et~al.}(1987)\citenamefont{Bak, Tang, and
  Wiesenfeld}}]{bak87}
\bibinfo{author}{\bibfnamefont{P.}~\bibnamefont{Bak}},
  \bibinfo{author}{\bibfnamefont{C.}~\bibnamefont{Tang}}, \bibnamefont{and}
  \bibinfo{author}{\bibfnamefont{K.}~\bibnamefont{Wiesenfeld}},
  \bibinfo{journal}{Phys.\ Rev.\ Lett.} \textbf{\bibinfo{volume}{59}},
  \bibinfo{pages}{381} (\bibinfo{year}{1987}).

\bibitem[{\citenamefont{Hwa and Kardar}(1989)}]{hwa89}
\bibinfo{author}{\bibfnamefont{T.}~\bibnamefont{Hwa}} \bibnamefont{and}
  \bibinfo{author}{\bibfnamefont{M.}~\bibnamefont{Kardar}},
  \bibinfo{journal}{Phys.\ Rev.\ Lett.} \textbf{\bibinfo{volume}{62}},
  \bibinfo{pages}{1813} (\bibinfo{year}{1989}).

\bibitem[{\citenamefont{Grinstein et~al.}(1990)\citenamefont{Grinstein, Lee,
  and Sachdev}}]{gls90}
\bibinfo{author}{\bibfnamefont{G.}~\bibnamefont{Grinstein}},
  \bibinfo{author}{\bibfnamefont{D.~H.} \bibnamefont{Lee}}, \bibnamefont{and}
  \bibinfo{author}{\bibfnamefont{S.}~\bibnamefont{Sachdev}},
  \bibinfo{journal}{Phys.\ Rev.\ Lett.} \textbf{\bibinfo{volume}{64}},
  \bibinfo{pages}{1927} (\bibinfo{year}{1990}).

\bibitem[{\citenamefont{Lise and Paczuski}(2001{\natexlab{a}})}]{lis01a}
\bibinfo{author}{\bibfnamefont{S.}~\bibnamefont{Lise}} \bibnamefont{and}
  \bibinfo{author}{\bibfnamefont{M.}~\bibnamefont{Paczuski}},
  \bibinfo{journal}{Phys.\ Rev.\ E} \textbf{\bibinfo{volume}{64}},
  \bibinfo{pages}{046111} (\bibinfo{year}{2001}{\natexlab{a}}).

\bibitem[{\citenamefont{Lise and Paczuski}(2001{\natexlab{b}})}]{lis01}
\bibinfo{author}{\bibfnamefont{S.}~\bibnamefont{Lise}} \bibnamefont{and}
  \bibinfo{author}{\bibfnamefont{M.}~\bibnamefont{Paczuski}},
  \bibinfo{journal}{Phys.\ Rev.\ E} \textbf{\bibinfo{volume}{63}},
  \bibinfo{pages}{036111} (\bibinfo{year}{2001}{\natexlab{b}}).

\bibitem[{\citenamefont{Miller and Boulter}(2003)}]{mb03a}
\bibinfo{author}{\bibfnamefont{G.}~\bibnamefont{Miller}} \bibnamefont{and}
  \bibinfo{author}{\bibfnamefont{C.}~\bibnamefont{Boulter}},
  \bibinfo{journal}{Phys.\ Rev.\ E} \textbf{\bibinfo{volume}{68}},
  \bibinfo{pages}{056108} (\bibinfo{year}{2003}).

\bibitem[{\citenamefont{de~Carvalho and Prado}(2000)}]{car00}
\bibinfo{author}{\bibfnamefont{J.}~\bibnamefont{de~Carvalho}} \bibnamefont{and}
  \bibinfo{author}{\bibfnamefont{C.}~\bibnamefont{Prado}},
  \bibinfo{journal}{Phys.\ Rev.\ Lett.} \textbf{\bibinfo{volume}{84}},
  \bibinfo{pages}{4006} (\bibinfo{year}{2000}).

\bibitem[{\citenamefont{Miller and Boulter}(2002)}]{mb02}
\bibinfo{author}{\bibfnamefont{G.}~\bibnamefont{Miller}} \bibnamefont{and}
  \bibinfo{author}{\bibfnamefont{C.}~\bibnamefont{Boulter}},
  \bibinfo{journal}{Phys.\ Rev.\ E} \textbf{\bibinfo{volume}{66}},
  \bibinfo{pages}{016123} (\bibinfo{year}{2002}).

\bibitem[{\citenamefont{Middleton and Tang}(1995)}]{mid95}
\bibinfo{author}{\bibfnamefont{A.}~\bibnamefont{Middleton}} \bibnamefont{and}
  \bibinfo{author}{\bibfnamefont{C.}~\bibnamefont{Tang}},
  \bibinfo{journal}{Phys.\ Rev.\ Lett.} \textbf{\bibinfo{volume}{74}},
  \bibinfo{pages}{742} (\bibinfo{year}{1995}).

\bibitem[{\citenamefont{Grassberger}(1994)}]{gras94}
\bibinfo{author}{\bibfnamefont{P.}~\bibnamefont{Grassberger}},
  \bibinfo{journal}{Phys.\ Rev.\ E} \textbf{\bibinfo{volume}{49}},
  \bibinfo{pages}{2436} (\bibinfo{year}{1994}).

\bibitem[{\citenamefont{P\'erez et~al.}(1996)\citenamefont{P\'erez, Corral,
  D\'iaz-Guilera, Christensen, and Arenas}}]{per96}
\bibinfo{author}{\bibfnamefont{C.}~\bibnamefont{P\'erez}},
  \bibinfo{author}{\bibfnamefont{A.}~\bibnamefont{Corral}},
  \bibinfo{author}{\bibfnamefont{A.}~\bibnamefont{D\'iaz-Guilera}},
  \bibinfo{author}{\bibfnamefont{K.}~\bibnamefont{Christensen}},
  \bibnamefont{and} \bibinfo{author}{\bibfnamefont{A.}~\bibnamefont{Arenas}},
  \bibinfo{journal}{Int. J. Mod. Phys. B} \textbf{\bibinfo{volume}{10}},
  \bibinfo{pages}{1111} (\bibinfo{year}{1996}).

\bibitem[{\citenamefont{Mousseau}(1996)}]{mou96}
\bibinfo{author}{\bibfnamefont{N.}~\bibnamefont{Mousseau}},
  \bibinfo{journal}{Phys.\ Rev.\ Lett.} \textbf{\bibinfo{volume}{77}},
  \bibinfo{pages}{968} (\bibinfo{year}{1996}).

\bibitem[{\citenamefont{J\'anosi and Kert\'esz}(1993)}]{jan93}
\bibinfo{author}{\bibfnamefont{I.}~\bibnamefont{J\'anosi}} \bibnamefont{and}
  \bibinfo{author}{\bibfnamefont{J.}~\bibnamefont{Kert\'esz}},
  \bibinfo{journal}{Physica A} \textbf{\bibinfo{volume}{200}},
  \bibinfo{pages}{179} (\bibinfo{year}{1993}).

\bibitem[{\citenamefont{Ceva}(1995)}]{ceva95}
\bibinfo{author}{\bibfnamefont{H.}~\bibnamefont{Ceva}},
  \bibinfo{journal}{Phys.\ Rev.\ E} \textbf{\bibinfo{volume}{52}},
  \bibinfo{pages}{154} (\bibinfo{year}{1995}).

\bibitem[{\citenamefont{Drossel}(2002)}]{dro02}
\bibinfo{author}{\bibfnamefont{B.}~\bibnamefont{Drossel}},
  \bibinfo{journal}{Phys.\ Rev.\ Lett.} \textbf{\bibinfo{volume}{89}},
  \bibinfo{pages}{238701} (\bibinfo{year}{2002}).

\bibitem[{\citenamefont{Hergarten and Neugebauer}(2002)}]{her02}
\bibinfo{author}{\bibfnamefont{S.}~\bibnamefont{Hergarten}} \bibnamefont{and}
  \bibinfo{author}{\bibfnamefont{H.}~\bibnamefont{Neugebauer}},
  \bibinfo{journal}{Phys.\ Rev.\ Lett.} \textbf{\bibinfo{volume}{88}},
  \bibinfo{pages}{238501} (\bibinfo{year}{2002}).

\bibitem[{\citenamefont{A.~Helmstetter and Sornette}(2004)}]{hhs04}
\bibinfo{author}{\bibfnamefont{S.~H.} \bibnamefont{A.~Helmstetter}}
  \bibnamefont{and} \bibinfo{author}{\bibfnamefont{D.}~\bibnamefont{Sornette}},
  \bibinfo{journal}{Phys.\ Rev.\ E} \textbf{\bibinfo{volume}{70}},
  \bibinfo{pages}{046120} (\bibinfo{year}{2004}).

\bibitem[{\citenamefont{Ramos et~al.}(2006)\citenamefont{Ramos, Altshuler, and
  M{\r a}l{\o}y}}]{ram06}
\bibinfo{author}{\bibfnamefont{O.}~\bibnamefont{Ramos}},
  \bibinfo{author}{\bibfnamefont{E.}~\bibnamefont{Altshuler}},
  \bibnamefont{and} \bibinfo{author}{\bibfnamefont{K.}~\bibnamefont{M{\r
  a}l{\o}y}}, \bibinfo{journal}{Phys.\ Rev.\ Lett.}
  \textbf{\bibinfo{volume}{96}}, \bibinfo{pages}{098501}
  (\bibinfo{year}{2006}).

\bibitem[{\citenamefont{Drossel and Wissel}(2005)}]{dro05}
\bibinfo{author}{\bibfnamefont{B.}~\bibnamefont{Drossel}} \bibnamefont{and}
  \bibinfo{author}{\bibfnamefont{F.}~\bibnamefont{Wissel}},
  \bibinfo{journal}{New Journal of Physics} \textbf{\bibinfo{volume}{7}},
  \bibinfo{pages}{5} (\bibinfo{year}{2005}).

\bibitem[{\citenamefont{Lise}(2002)}]{lis02}
\bibinfo{author}{\bibfnamefont{S.}~\bibnamefont{Lise}},
  \bibinfo{journal}{J.Phys. A} \textbf{\bibinfo{volume}{35}},
  \bibinfo{pages}{4641} (\bibinfo{year}{2002}).

\bibitem[{\citenamefont{Ceva}(1998)}]{ceva98}
\bibinfo{author}{\bibfnamefont{H.}~\bibnamefont{Ceva}},
  \bibinfo{journal}{Phys.\ Lett.\ A} \textbf{\bibinfo{volume}{245}},
  \bibinfo{pages}{413} (\bibinfo{year}{1998}).

\bibitem[{\citenamefont{Bottani and Delamotte}(1997)}]{bot97}
\bibinfo{author}{\bibfnamefont{S.}~\bibnamefont{Bottani}} \bibnamefont{and}
  \bibinfo{author}{\bibfnamefont{B.}~\bibnamefont{Delamotte}},
  \bibinfo{journal}{Physica D} \textbf{\bibinfo{volume}{103}},
  \bibinfo{pages}{430} (\bibinfo{year}{1997}).

\bibitem[{\citenamefont{Christensen et~al.}(2001)\citenamefont{Christensen,
  Hamon, Jensen, and Lise}}]{ch00}
\bibinfo{author}{\bibfnamefont{K.}~\bibnamefont{Christensen}},
  \bibinfo{author}{\bibfnamefont{D.}~\bibnamefont{Hamon}},
  \bibinfo{author}{\bibfnamefont{H.}~\bibnamefont{Jensen}}, \bibnamefont{and}
  \bibinfo{author}{\bibfnamefont{S.}~\bibnamefont{Lise}},
  \bibinfo{journal}{Phys.\ Rev.\ Lett.} \textbf{\bibinfo{volume}{87}},
  \bibinfo{pages}{039801} (\bibinfo{year}{2001}).

\bibitem[{\citenamefont{de~Carvalho and Prado}(2001)}]{cp01}
\bibinfo{author}{\bibfnamefont{J.}~\bibnamefont{de~Carvalho}} \bibnamefont{and}
  \bibinfo{author}{\bibfnamefont{C.}~\bibnamefont{Prado}},
  \bibinfo{journal}{Phys.\ Rev.\ Lett.} \textbf{\bibinfo{volume}{87}},
  \bibinfo{pages}{039802} (\bibinfo{year}{2001}).

\bibitem[{\citenamefont{Lise and Jensen}(1996)}]{lj96}
\bibinfo{author}{\bibfnamefont{S.}~\bibnamefont{Lise}} \bibnamefont{and}
  \bibinfo{author}{\bibfnamefont{H.}~\bibnamefont{Jensen}},
  \bibinfo{journal}{Phys.\ Rev.\ Lett.} \textbf{\bibinfo{volume}{76}},
  \bibinfo{pages}{2326} (\bibinfo{year}{1996}).

\bibitem[{\citenamefont{Chabanol and Hakim}(1997)}]{ch97}
\bibinfo{author}{\bibfnamefont{M.}~\bibnamefont{Chabanol}} \bibnamefont{and}
  \bibinfo{author}{\bibfnamefont{V.}~\bibnamefont{Hakim}},
  \bibinfo{journal}{Phys.\ Rev.\ E} \textbf{\bibinfo{volume}{56}},
  \bibinfo{pages}{R2343} (\bibinfo{year}{1997}).

\bibitem[{\citenamefont{Broeker and Grassberger}(1997)}]{bg97}
\bibinfo{author}{\bibfnamefont{H.}~\bibnamefont{Broeker}} \bibnamefont{and}
  \bibinfo{author}{\bibfnamefont{P.}~\bibnamefont{Grassberger}},
  \bibinfo{journal}{Phys.\ Rev.\ E} \textbf{\bibinfo{volume}{56}},
  \bibinfo{pages}{3944} (\bibinfo{year}{1997}).

\bibitem[{\citenamefont{Pinho et~al.}(1998)\citenamefont{Pinho, Prado, and
  Kinouchi}}]{ppk98}
\bibinfo{author}{\bibfnamefont{S.}~\bibnamefont{Pinho}},
  \bibinfo{author}{\bibfnamefont{C.}~\bibnamefont{Prado}}, \bibnamefont{and}
  \bibinfo{author}{\bibfnamefont{O.}~\bibnamefont{Kinouchi}},
  \bibinfo{journal}{Physica A} \textbf{\bibinfo{volume}{257}},
  \bibinfo{pages}{488} (\bibinfo{year}{1998}).

\bibitem[{\citenamefont{Burridge and Knopoff}(1967)}]{bur67}
\bibinfo{author}{\bibfnamefont{R.}~\bibnamefont{Burridge}} \bibnamefont{and}
  \bibinfo{author}{\bibfnamefont{L.}~\bibnamefont{Knopoff}},
  \bibinfo{journal}{Bull.\ Seismol.\ Soc.\ Am.} \textbf{\bibinfo{volume}{57}},
  \bibinfo{pages}{341} (\bibinfo{year}{1967}).

\bibitem[{\citenamefont{Miller and Boulter}(2004)}]{mb03}
\bibinfo{author}{\bibfnamefont{G.}~\bibnamefont{Miller}} \bibnamefont{and}
  \bibinfo{author}{\bibfnamefont{C.}~\bibnamefont{Boulter}},
  \bibinfo{journal}{Phys.\ Rev.\ E} \textbf{\bibinfo{volume}{67}},
  \bibinfo{pages}{046114} (\bibinfo{year}{2004}).

\bibitem[{\citenamefont{Christensen et~al.}(1992)\citenamefont{Christensen,
  Olami, and Bak}}]{cob92}
\bibinfo{author}{\bibfnamefont{K.}~\bibnamefont{Christensen}},
  \bibinfo{author}{\bibfnamefont{Z.}~\bibnamefont{Olami}}, \bibnamefont{and}
  \bibinfo{author}{\bibfnamefont{P.}~\bibnamefont{Bak}},
  \bibinfo{journal}{Phys.\ Rev.\ Lett.} \textbf{\bibinfo{volume}{68}},
  \bibinfo{pages}{2417} (\bibinfo{year}{1992}).

\bibitem[{\citenamefont{Olami and Christensen}(1992)}]{oc92}
\bibinfo{author}{\bibfnamefont{Z.}~\bibnamefont{Olami}} \bibnamefont{and}
  \bibinfo{author}{\bibfnamefont{K.}~\bibnamefont{Christensen}},
  \bibinfo{journal}{Phys.\ Rev.\ A.} \textbf{\bibinfo{volume}{46}},
  \bibinfo{pages}{R1720} (\bibinfo{year}{1992}).

\bibitem[{\citenamefont{Christensen and Olami}(1992)}]{co92}
\bibinfo{author}{\bibfnamefont{K.}~\bibnamefont{Christensen}} \bibnamefont{and}
  \bibinfo{author}{\bibfnamefont{Z.}~\bibnamefont{Olami}},
  \bibinfo{journal}{Phys.\ Rev.\ A} \textbf{\bibinfo{volume}{46}},
  \bibinfo{pages}{1829} (\bibinfo{year}{1992}).

\bibitem[{\citenamefont{Pepke and Carlson}(1994)}]{pc94}
\bibinfo{author}{\bibfnamefont{S.}~\bibnamefont{Pepke}} \bibnamefont{and}
  \bibinfo{author}{\bibfnamefont{J.}~\bibnamefont{Carlson}},
  \bibinfo{journal}{Phys.\ Rev.\ E} \textbf{\bibinfo{volume}{40}},
  \bibinfo{pages}{234} (\bibinfo{year}{1994}).

\end{thebibliography}

\end{document}